# Multi-Hill Strategy in Metadynamics for Interstitial Diffusion in Crystals

Kazuaki Toyoura*

*Department of Materials Science and Engineering, Kyoto University, Kyoto 606-8501, Japan*

* toyoura.kazuaki.5r@kyoto-u.ac.jp

**Abstract**

We propose an efficient and general strategy of metadynamics (MetaD) for investigating interstitial diffusion in a crystal by exploiting crystallographic symmetry. Assuming complete ignorance of the diffusion phenomenon of interest, the three-dimensional coordinates of the interstitial atom with the periodic boundaries are chosen as the collective variables (CVs). Multiple potential hills are simultaneously deposited at all crystallographically-equivalent positions on the free energy surface (FES) defined in the CV space. As a result, the proposed multi-hill strategy highly accelerates atomic jumps in comparison with the single-hill strategy in the conventional MetaD. The key features are that the FES estimated from the final bias potential is exactly satisfied with the symmetry of the host crystal and that all elementary processes of interstitial diffusion are obtained by the single MetaD simulation without any prior knowledge on the diffusion mechanism. The high efficiency and efficacy of the multi-hill strategy are demonstrated, taking the proton diffusion in barium zirconate with the cubic perovskite structure as a model case.

# I. INTRODUCTION

Molecular dynamics (MD) is an excellent tool for understanding diffusion phenomena in solids at the atomic level, which tracks the time evolution of a given system according to Newton's equation of motion. However, the simulation time scale is limited, e.g., 1 ns at most in the case of first-principles molecular dynamics (FPMD) simulations. Therefore, rare events with a longer time scale than the simulation time scale cannot be treated by conventional MD simulations. In addition, quantitative evaluation of atomic jump frequencies requires observation of multiple jumps to obtain statistical averages.

Various techniques accelerating rare events in MD simulations have been reported so far, such as umbrella sampling [1,2], hyperdynamics [3,4], metadynamics [5-8], and variational enhanced sampling [9]. In these techniques, a free energy surface (FES) is considered in the subspace of configuration space defined by a few given collective variables (CVs) describing the physical phenomenon of interest. An external bias potential is added on the low-dimensional FES to accelerate rare events in MD simulations. Of importance is how to choose suitable CVs for describing the physical phenomenon of interest [8,10,11]. There are a variety of CV candidates, e.g., atomic positions, interatomic distances, bond angles, torsional angles, interbond distances, components of simulation cell (lattice parameters), reaction paths, and their combinations. We should choose the necessary and sufficient CVs because the lower-dimensional CV space leads to the more effective acceleration.

In the present study, we focus on metadynamics (MetaD) as one of the most widely-used adaptive bias simulation methods, and propose an efficient and general utilization of MetaD for investigating interstitial diffusion in a crystal. Assuming complete ignorance of the diffusion phenomenon of interest, the three-dimensional fractional coordinates **x** with the periodic boundaries are employed as the CVs. In the conventional MetaD method, a single small Gaussian hill is iteratively deposited on the FES at a given time interval at the current position in the CV space to gradually fill the free energy basins (*single-hill strategy*). By contrast, the proposed method takes the strategy that multiple Gaussian hills are deposited at the same time by exploiting the symmetry of host crystals to enhance the basin-filling efficiency (*multi-hill strategy*). We here demonstrate how the multi-hill strategy works efficiently and effectively in comparison with the single-hill strategy, taking the proton diffusion in barium zirconate ($BaZrO_3$) with the cubic perovskite structure as a model case.

## II. COMPUTATIONAL METHODOLOGY

**A. Proposed method**

In the conventional MetaD method, the bias potential $U_n^{\text{bias}}(\mathbf{s})$ after $n$-times deposition of Gaussian hills is given by

$$U_n^{\text{bias}}(\mathbf{s}) = U_{n-1}^{\text{bias}}(\mathbf{s}) + h_n \exp\left(-\frac{(\mathbf{s}-\mathbf{s}_n)^T \Sigma^{-1} (\mathbf{s}-\mathbf{s}_n)}{2}\right) \approx U_{n-1}^{\text{bias}}(\mathbf{s}) + h_n \exp\left(-\frac{\|\mathbf{s}-\mathbf{s}_n\|^2}{2\sigma^2}\right), \quad (1)$$

where $h_n$ and $\Sigma$ are the height and covariance matrix of the $n$th Gaussian hill, respectively. **s** is the vector of CVs, and $\mathbf{s}_n$ is the current position in the CV space when the $n$th Gaussian hill is added. When $\Sigma$ is the diagonal matrix with all elements equal, the Gaussian hills can be rewritten simply as the rightmost side of Eq. (1) using the common standard deviation $\sigma$. $h_n$ is set to be constant ($h_0$) in the original MetaD, while $h_n$ is rescaled according to the bias potential at the current position $\mathbf{s}_n$ in well-tempered MetaD (WT-MetaD) as follows:

$$h_n = h_0 \exp\left(-\frac{V_{n-1}(\mathbf{s}_n)}{k_B \Delta T}\right), \tag{2}$$

where $k_B$ is the Boltzmann constant and $\Delta T$ is a temperature-valued positive parameter determining the decay rate of the Gaussian hill height. The scaling exponential factor is unity at $\Delta T = \infty$, equivalent to the original MetaD. After all basins on the FES are sufficiently filled up by the deposited Gaussian hills, the bias potential $U_{n_{\max}}^{\text{bias}}(\mathbf{s})$ can be converted into the FES $F(\mathbf{s})$ as follows [6]:

$$F(\mathbf{s}) = -(1 + T/\Delta T)\, U_{n_{\max}}^{\text{bias}}(\mathbf{s}) + const., \tag{3}$$

where $T$ is the simulation temperature.

In the proposed method, the three-dimensional fractional coordinates $\mathbf{x}$ of the interstitial atom with the periodic boundaries are defined as CVs. The defined CV space corresponds to the entire simulation cell, which is too large to be filled up by depositing small Gaussian hills one by one. Therefore, the proposed method takes the strategy that multiple Gaussian hills are simultaneously deposited at all crystallographically equivalent positions in the simulation cell. In this strategy, the bias potential $U_n^{\text{bias}}(\mathbf{x})$ is given by

$$U_n^{\text{bias}}(\mathbf{x}) = U_{n-1}^{\text{bias}}(\mathbf{x}) + \sum_k h_n \exp\left(-\frac{(\mathbf{x} - \mathbf{T}_k \cdot \mathbf{x}_n)^T \Sigma^{-1}(\mathbf{x} - \mathbf{T}_k \cdot \mathbf{x}_n)}{2}\right) \approx U_{n-1}^{\text{bias}}(\mathbf{x}) + \sum_k h_n \exp\left(-\frac{\|\mathbf{x} - \mathbf{T}_k \cdot \mathbf{x}_n\|^2}{2\sigma^2}\right), \tag{4}$$

where $\mathbf{T}_k$ is the symmetry operation that generates the $k$th point equivalent to the current position $\mathbf{x}_n$ of the interstitial atom.

**B. Model system and computational conditions**

The proton diffusion in the model system of BaZrO$_3$ has been investigated intensively using theoretical calculations [12-16]. Figure 1 shows the proton interstitial sites (white small spheres) and two types of migration paths (black and white lines). A proton resides around an oxide ion and migrates by repeating rotation around an oxide ion (black lines) and hopping between adjacent oxide ions (white lines). A supercell consisting of 2×2×2 BaZrO$_3$ unit cells including a single proton was employed as the simulation cell. The goal of the demonstration problem is that the proton FESs around the rotation and hopping paths are evaluated accurately to estimate the proton jump frequencies. WT-MetaD simulations were carried out using the proposed multi-hill strategy, and those under the conventional single-hill strategy were also carried out for comparison. Only for the single-hill strategy, the mobile region of a proton was limited within the two basin regions corresponding to the initial and final states for each of the rotation and hopping paths. Specifically, the polyhedral region composed of the gray and blue tetrahedra in Fig. 1 was surrounded by elastic collision walls for the rotation path, while the region composed of the gray and green tetrahedra was surrounded by walls for the hopping path.

All simulations were based on first-principles calculations using the projector augmented wave (PAW) method [17,18] implemented in the Vienna Ab initio Simulation Package (VASP) code [19-21]. The generalized gradient approximation parameterized by Perdew, Burke, and Ernzerhof (PBE_GGA) was used for the exchange-correlation term [22]. The 5$s$, 5$p$, 6$s$, and 5$d$ orbitals for Ba, 4*s*, 4*p*, 5*s*, and 4*d* for Zr and Y, 2*s* and 2*p* for O, and 1*s* for H were treated as valence states in the PAW potentials. The plane wave cutoff energy was set to 400 eV. Brillouin-zone sampling was performed using a 2 × 2 × 2 grid for the supercell consisting of 2 × 2 × 2 unit cells. The *NVT* ensemble was employed with the constant temperature controlled by the Nóse-Hoover thermostat [23]. The time step was set to $1 \times 10^{-15}$ s (1 fs) and the total simulation steps were 150000 and 600000 steps for the multi-hill and single-hill strategies, respectively. Pre-simulations were preliminarily performed for 10000 steps as thermal equilibration. Gaussian hills were deposited at every 50 MD steps, i.e., 3000-times and 12000-times hill depositions under the multi-hill and single-hill strategies, respectively. The height and standard deviation of the Gaussian hills ($h_0$ and $\sigma$) and the scaling parameter ($k_B \Delta T$) were set to 0.02 eV, 0.01, and 0.1 eV, respectively.

**C. Atomic jump frequencies**

Proton jump frequencies were estimated from the obtained FESs based on the transition state theory [24]. Within classical statistical mechanics, the jump frequency $\nu$ is given by

$$\nu = \sqrt{\frac{k_B T}{2\pi M}} \frac{\iint_S \exp(-\frac{F(\mathbf{x})}{k_B T}) dS}{\iiint_V \exp(-\frac{F(\mathbf{x})}{k_B T}) dV}, \tag{5}$$

where $M$ is the mass of diffusion species (protons). The denominator and numerator are the volume and area integrations, respectively, where the integral ranges are the single basin corresponding to the initial state and the saddle surface separating the initial and final states, respectively. For the proton rotation and hopping in $BaZrO_3$, the volume and area integrations were performed within a tetrahedron shown in Fig. 1 and on the face shared by two adjacent tetrahedra, respectively.

The jump frequencies under the classical harmonic approximation were also estimated to examine the accuracy of the jump frequencies obtained from the WT-MetaD simulations. Under the classical harmonic approximation, the jump frequency is given by

$$\nu = \nu_{0,\text{harm}} \exp(-\frac{\Delta E^{\text{mig}}}{k_B T}) = \frac{\prod_{i=1}^{3N-3} \nu_i^{\text{initial}}}{\prod_{i=2}^{3N-3} \nu_i^{\text{saddle}}} \exp(-\frac{\Delta E^{\text{mig}}}{k_B T}), \qquad (6)$$

where $\nu_{0,\text{harm}}$ is the vibrational prefactor expressed by the eigenfrequencies of lattice vibrations in the initial and saddle-point states, $\nu_i^{\text{initial}}$ and $\nu_i^{\text{saddle}}$, respectively, $\Delta E^{\text{mig}}$ is the potential barrier, and $N$ is the number of atoms in a given system. The eigenfrequencies in the initial state and the saddle-point states for proton rotation and hopping in $BaZrO_3$ were estimated using the phonopy code [25] based on the small displacement method [26,27]. Figure 2 shows the estimated eigenfrequencies at the Γ point in the initial and saddle-point states for proton rotation and hopping. The single imaginary frequency in each of the saddle-point states corresponds to the local energy maximum along the migration coordinate, which is excluded in evaluating the jump frequencies according to Eq. (6).

**III. Results and Discussion**

**A. Hill deposition profiles and FESs**

Figure 3 shows the deposition profiles of Gaussian hills in the $BaZrO_3$ supercell with a single proton during the WT-MetaD simulations at 500 K. The cubic perovskite structure has space group $Pm\bar{3}m$ (221), in which the multiplicity of the general position is 48. Therefore, 384 Gaussian hills are deposited in the supercell consisting of 2 × 2 × 2 unit cells at each deposition step (every 50 MD steps) under the multi-hill strategy. Under the single-hill strategy, the two deposition profiles corresponding to the proton rotation and hopping are shown in the figure, for which proton migration was limited within the two adjacent basins by elastic collision walls. Note that such limitation of proton migration is possible only by exploiting the prior knowledge shown in Fig. 1, i.e., infeasible before evaluating the entire FES. By contrast, the multi-hill strategy does not require such prior knowledge on the FES, which uses only the symmetry of the host crystal. As a result, the deposited hills are distributed according to the crystallographic symmetry over the entire simulation cell, which are denser than those under the single-hill strategy.

Figure 4(a) shows the cross-sections of the estimated FESs at 500 K on the proton rotation and hopping planes after 3000 and 12000 deposition steps (150000 and 600000 MD steps) under the multi-hill and single-hill strategies, respectively. The free energies are shown with reference to the most stable position in the CV space. As shown in the figure, the estimated FES exactly satisfies the symmetry of the host crystal under the multi-hill strategy, while only the two adjacent free energy minima exist under the single-hill strategy. Figure 4(b) shows the profiles of the cross-sections of the estimated FESs at 100, 500, 1000, 2000, and 3000 deposition steps (5000, 25000, 50000, 100000, and 150000 MD steps). At the earliest step, the basins around proton sites are shallow due to the insufficient deposition of Gaussian hills. As increasing the deposited hills, the basins become deeper gradually and the boundary regions between adjacent basins are also depressed, to enable estimating proton jump frequencies. The multi-hill strategy always creates the FESs according to the symmetry of the host crystal even at the early steps in principle, while the two basins corresponding to the initial and final states of the proton jump are different in depth and shape under the single-hill strategy, particularly at the early steps. Note that the multi-hill strategy can extract all elementary processes of interstitial diffusion from the FES obtained by the single simulation. In this model case, the proton rotation and hopping paths are identified at once by the optimal path finding on the finally-obtained FES.

**B. Jump frequencies**

Figure 5 shows the estimated jump frequencies of proton rotation and hopping at 500 K as a function of hill deposition steps. Under the multi-hill strategy, all rotation or hopping paths in the simulation cell are equivalently evaluated, while the forward and backward jumps between the two basins are not equivalent under the single-hill strategy. Therefore, the estimated jump frequencies are shown by the single lines (red lines with solid symbols) under the multi-hill strategy and by the two lines (green and blue lines with open symbols) under the single-hill strategy. The black broken lines denote the jump frequencies estimated under the classical harmonic approximation. Although the jump frequencies tend to be overestimated at the early steps due to the shallow free energy basins, the jump frequencies quickly converge to the broken lines under the multi-hill strategy. By contrast, the convergence of the two lines under the single-hill strategy is relatively slow. In addition, the two lines corresponding to the forward and backward jumps largely deviate from each other in the early steps, and they do not coincide with each other perfectly even in the later steps. In principle, the efficiency of the multi-hill strategy is 8 times higher than that of the single-hill strategy, because there are four crystallographically equivalent positions per basin in this model system. If comparing the two strategies under the same condition without elastic collision walls, the multi-hill strategy should be 384 times more efficient than the single-hill strategy.

The WT-MetaD simulations under the multi-hill strategy were performed at various temperatures in the range of 500 K and 2000 K, to estimate the jump frequencies for proton rotation and hopping. Figure 6 shows the estimated jump frequencies as a function of inverse temperature (red lines with solid symbols), in which those under the classical harmonic approximation are also shown by the broken lines for comparison. The apparent activation energies $Q$ and pre-exponential factors $\nu_{0,\mathrm{app}}$ for the jump frequencies are 0.17 eV and 11 THz for proton rotation, which are in reasonable agreement with the potential barrier $\Delta E^{\mathrm{mig}}$ (0.16 eV) and the vibrational prefactor $\nu_{0,\mathrm{harm}}$ (7 THz) under the harmonic approximation. However, $Q$ and $\nu_{0,\mathrm{app}}$ for proton hopping are 0.22 eV and 12 THz, respectively, which are comparatively different from the $\Delta E^{\mathrm{mig}}$ and $\nu_{0,\mathrm{harm}}$ under the harmonic approximation (0.26 eV and 30 THz). Although a possible origin of the discrepancy is anharmonicity, the estimation errors of jump frequencies by both methods are not negligible. In fact, the estimated jump frequencies by the WT-MetaD simulations have larger variation at lower temperatures, and the vibrational prefactors given by Eq. (6) are significantly affected by the errors of the estimated eigenfrequencies at the initial and saddle-point states, particularly by low-frequency modes.

## IV. CONCLUSIONS

We proposed the multi-hill strategy in MetaD simulations for evaluating interstitial diffusion in crystals by exploiting the crystallographic symmetry. Specifically, using the fractional coordinates of the interstitial with the periodic boundaries as the CVs, Gaussian hills are deposited at all positions equivalent to the current interstitial position. The higher symmetry of the simulation cell leads to the more Gaussian hill deposition, resulting in the more efficient simulation. The biggest advantage of the multi-hill strategy is that all elementary processes of interstitial diffusion in the crystal are obtained by the single MetaD simulation without any prior knowledge on the diffusion phenomenon of interest. The high efficiency and efficacy of the multi-hill strategy were demonstrated by taking WT-MetaD simulations for the proton diffusion in $BaZrO_3$ as a model case.

## ACKNOWLEDGMENTS

We gratefully acknowledge the insightful discussion with S. Takazawa. This work was partially supported by JSPS KAKENHI (grant numbers: 19H05787 and 20H02422).

**FIGURE CAPTIONS**

FIGURE 1. Proton sites and migration paths in 2 × 2 × 2 unit cells of $BaZrO_3$. The yellow, gray, and red spheres are Ba, Zr, and O ions, respectively, and the white small spheres are the proton sites. The black and white lines connecting adjacent proton sites denote proton rotation and hopping paths, respectively. There is the tetrahedral region corresponding to the basin on the FES around each proton site. The gray tetrahedron is adjacent to the blue and green tetrahedra with face sharing, and the two shared faces correspond to the saddle surfaces of rotation and hopping paths, respectively.

FIGURE 2. Estimated eigenfrequencies of lattice vibration at the initial state and the saddle-point states for proton rotation and hopping. The region below 0 THz indicates imaginary frequencies. The three red lines in each state correspond to the proton vibrations.

FIGURE 3. Hill deposition profiles in the $BaZrO_3$ supercell consisting of 2 × 2 × 2 unit cells with a single proton during the WT-MetaD simulations at 500 K under the multi-hill and single-hill strategies. The black dots are the positions of deposited hills at 100, 500, and 1000 deposition steps. The yellow surfaces denote the free energy isosurfaces at 0.5 and 0.4 eV vs. the most stable site under the multi-hill and single-hill strategies, respectively, after finishing the WT-MetaD simulations. Under the multi-hill strategy, 384 hills are deposited on the FES at each step.

FIGURE 4. (a) Cross-sections of the estimated FESs on the proton rotation and hopping planes after 3000 and 12000 deposition steps (150000 and 600000 MD steps) under the multi-hill and single-hill strategies. (b) Cross-sections of the estimated FESs at 100, 500, 1000, 2000, and 3000 deposition steps. The free energy is shown with reference to the most stable position at each step.

FIGURE 5. Estimated jump frequencies of (a) proton rotation and (b) proton hopping under the multi-hill and single-hill strategies as a function of the deposition step. The red lines with solid symbols correspond to the multi-hill strategy, while the green and blue lines with open symbols correspond to the forward and backward jumps under the single-hill strategy. The broken vertical lines are the jump frequencies estimated under the classical harmonic approximation by Eq. (6).

FIGURE 6. Temperature dependence of the estimated jump frequencies for (a) proton rotation and (b) proton hopping under the multi-hill strategy. The estimated jump frequencies under the classical harmonic approximation are also shown for comparison.

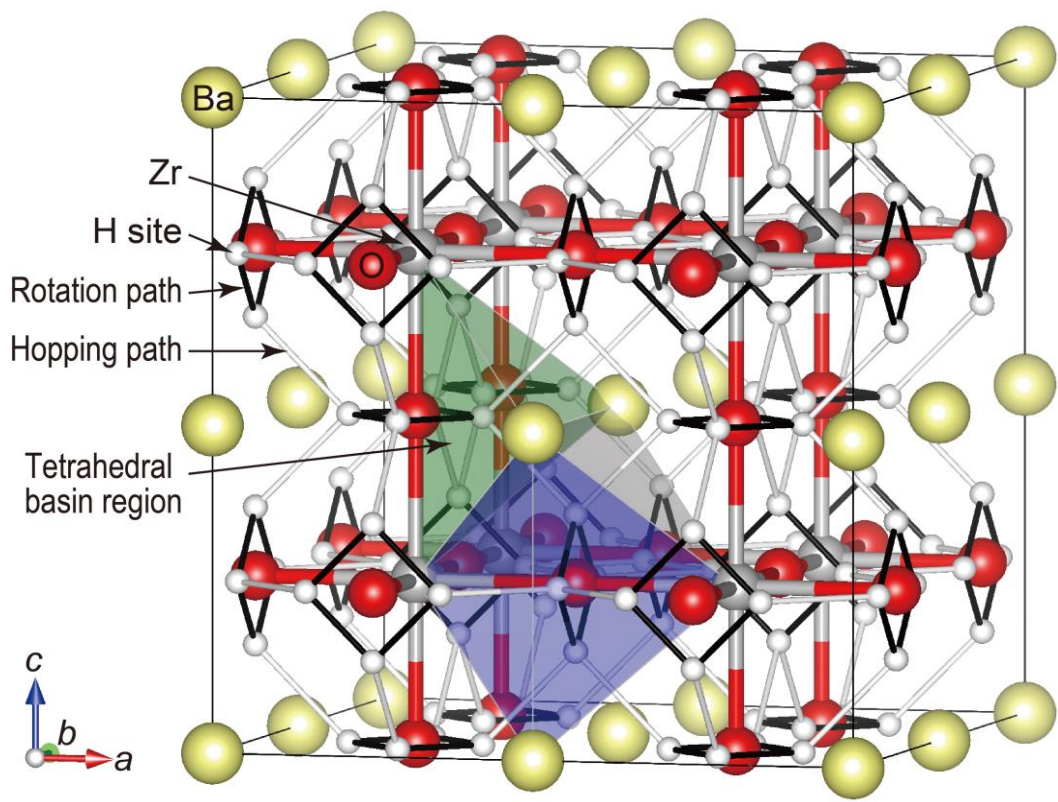

FIGURE 1. Proton sites and migration paths in 2 × 2 × 2 unit cells of $BaZrO_3$. The yellow, gray, and red spheres are Ba, Zr, and O ions, respectively, and the white small spheres are the proton sites. The black and white lines connecting adjacent proton sites denote proton rotation and hopping paths, respectively. There is the tetrahedral region corresponding to the basin on the FES around each proton site. The gray tetrahedron is adjacent to the blue and green tetrahedra with face sharing, and the two shared faces correspond to the saddle surfaces of rotation and hopping paths, respectively.

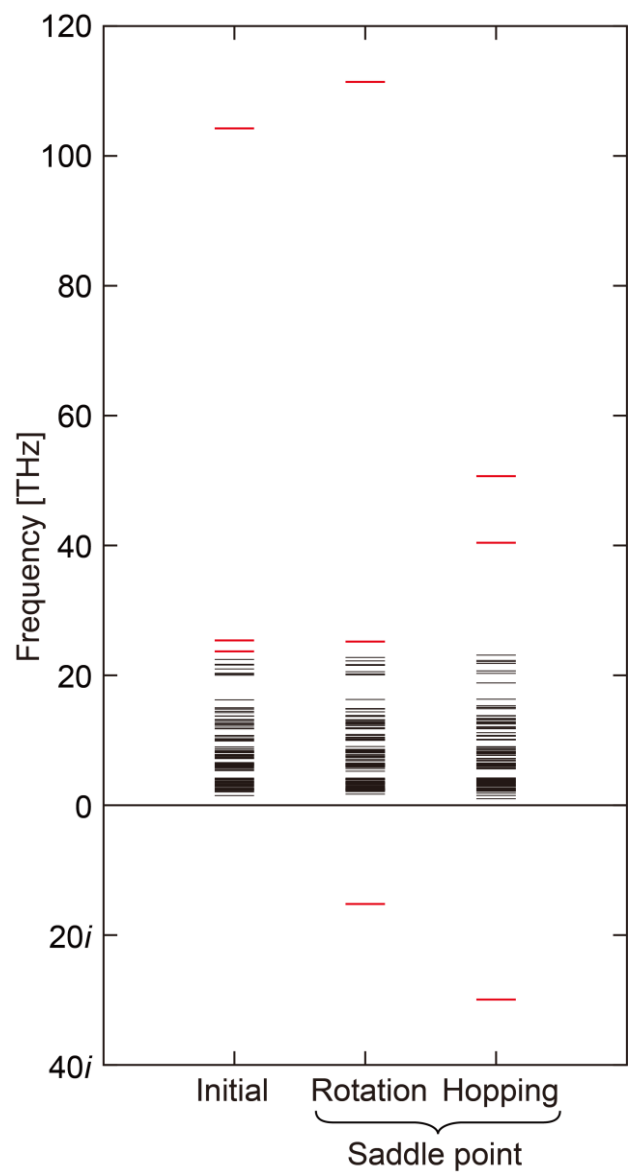

FIGURE 2. Estimated eigenfrequencies of lattice vibration at the initial state and the saddle-point states for proton rotation and hopping. The region below 0 THz indicates imaginary frequencies. The three red lines in each state correspond to the proton vibrations.

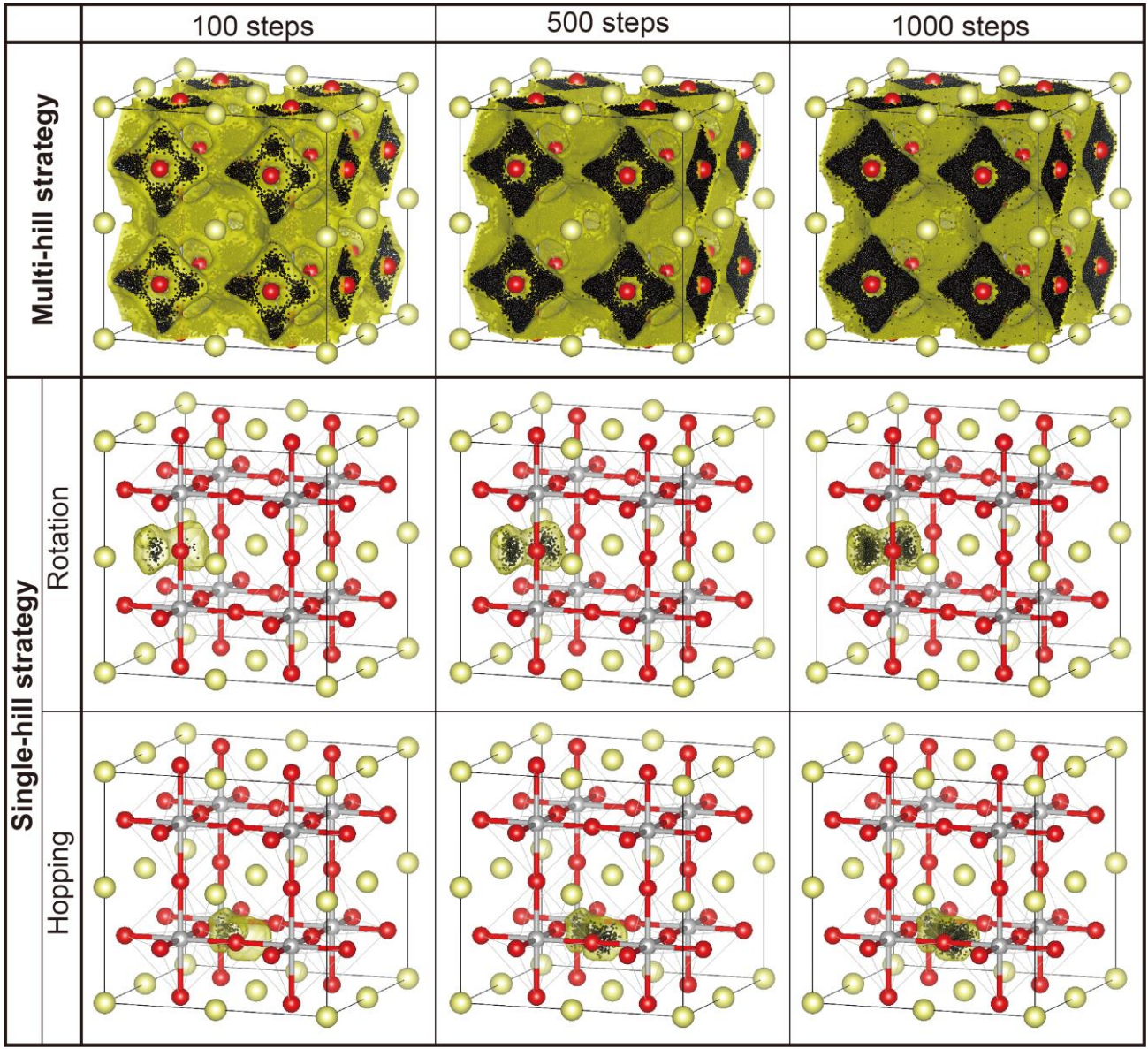

FIGURE 3. Hill deposition profiles in the BaZrO$_3$ supercell consisting of $2 \times 2 \times 2$ unit cells with a single proton during the WT-MetaD simulations at 500 K under the multi-hill and single-hill strategies. The black dots are the positions of deposited hills at 100, 500, and 1000 deposition steps. The yellow surfaces denote the free energy isosurfaces at 0.5 and 0.4 eV vs. the most stable site under the multi-hill and single-hill strategies, respectively, after finishing the WT-MetaD simulations. Under the multi-hill strategy, 384 hills are deposited on the FES at each step.

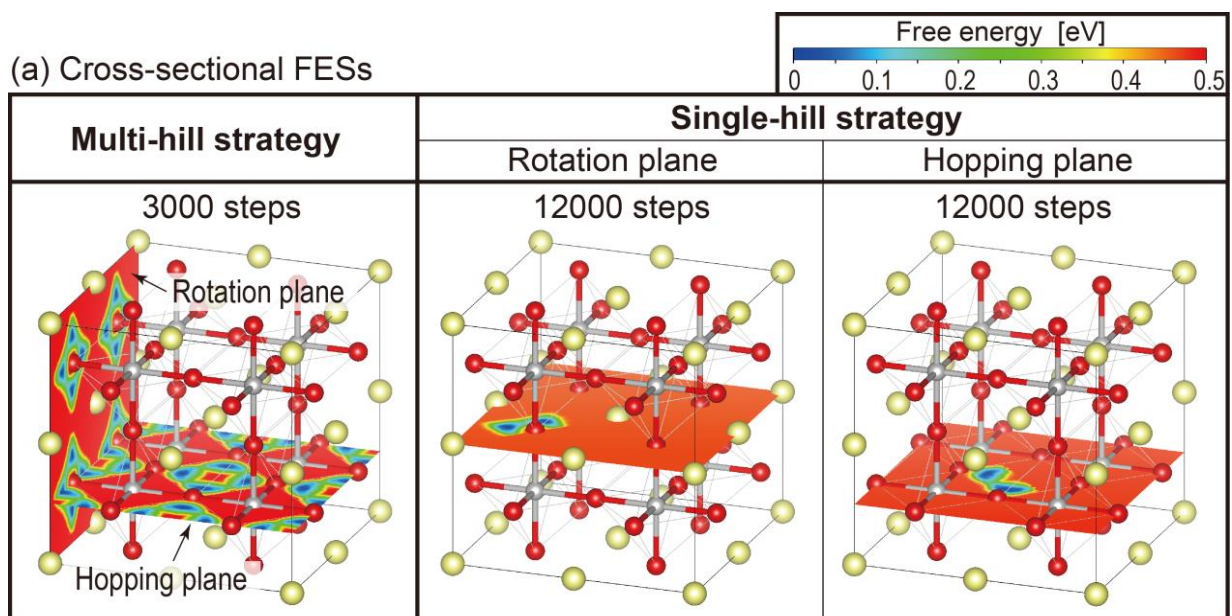

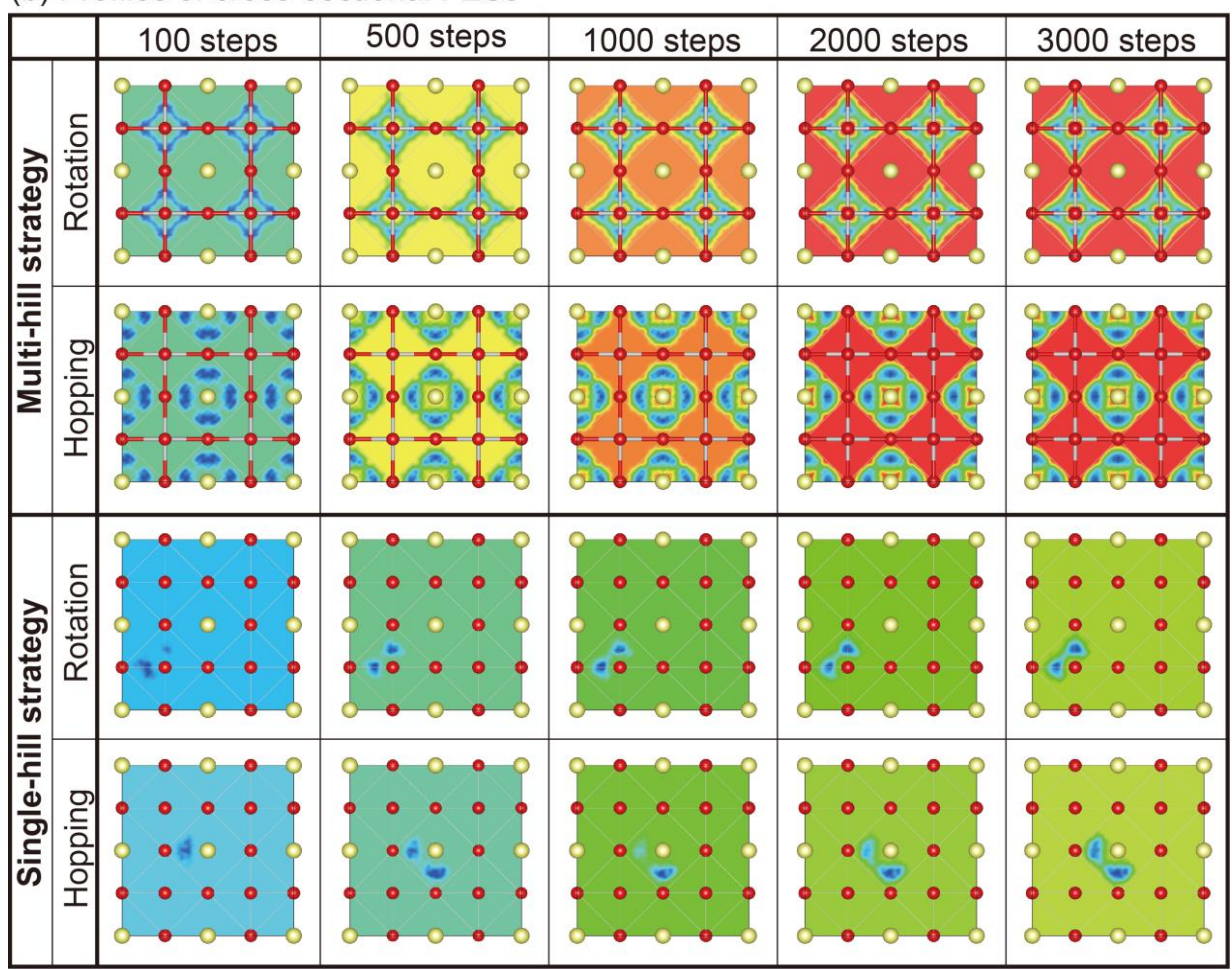

FIGURE 4. (a) Cross-sections of the estimated FESs on the proton rotation and hopping planes after 3000 and 12000 deposition steps (150000 and 600000 MD steps) under the multi-hill and single-hill strategies. (b) Cross-sections of the estimated FESs at 100, 500, 1000, 2000, and 3000 deposition steps. The free energy is shown with reference to the most stable position at each step.

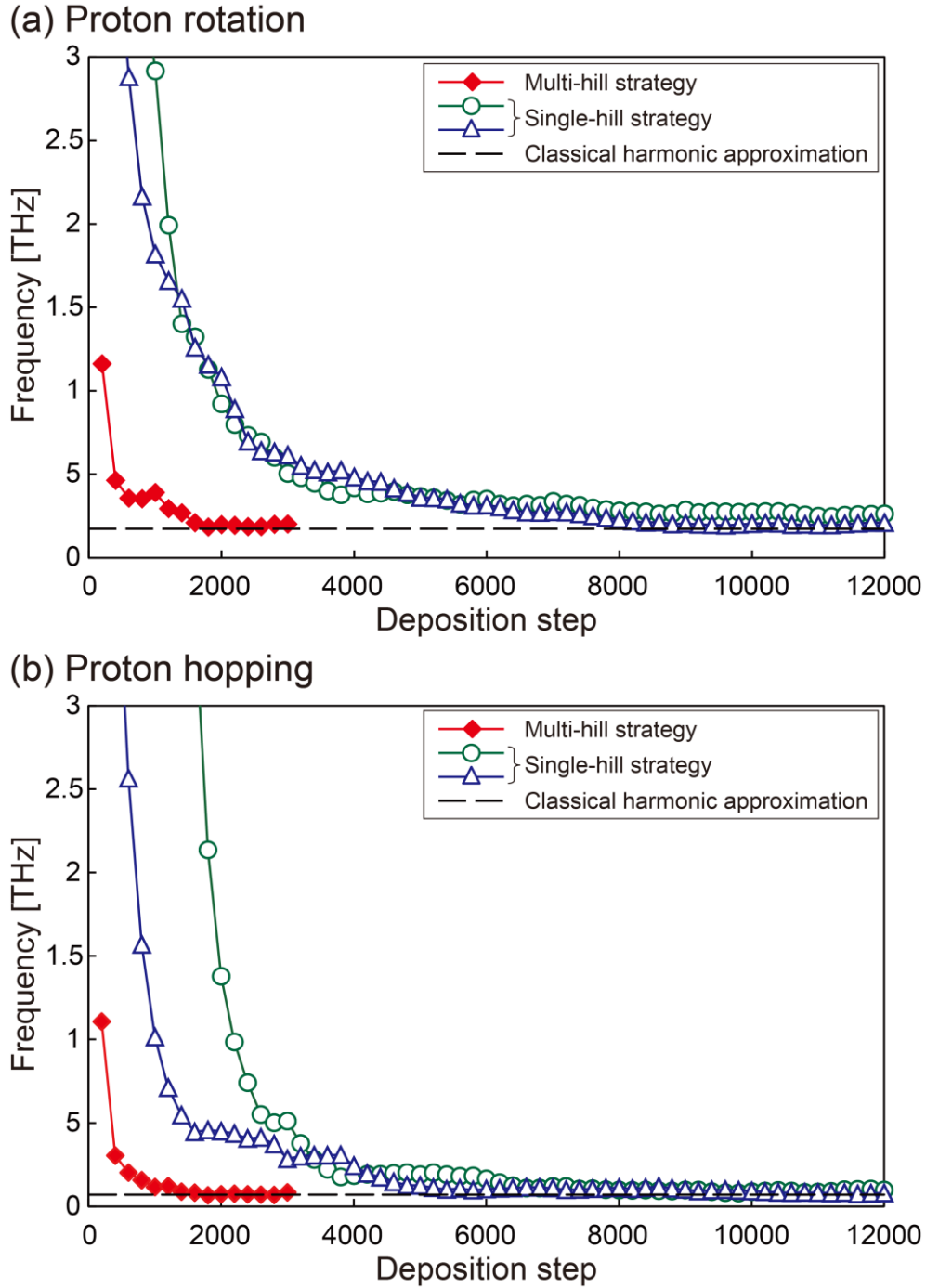

FIGURE 5. Estimated jump frequencies of (a) proton rotation and (b) proton hopping under the multi-hill and single-hill strategies as a function of the deposition step. The red lines with solid symbols correspond to the multi-hill strategy, while the green and blue lines with open symbols correspond to the forward and backward jumps under the single-hill strategy. The broken vertical lines are the jump frequencies estimated under the classical harmonic approximation by Eq. (6).

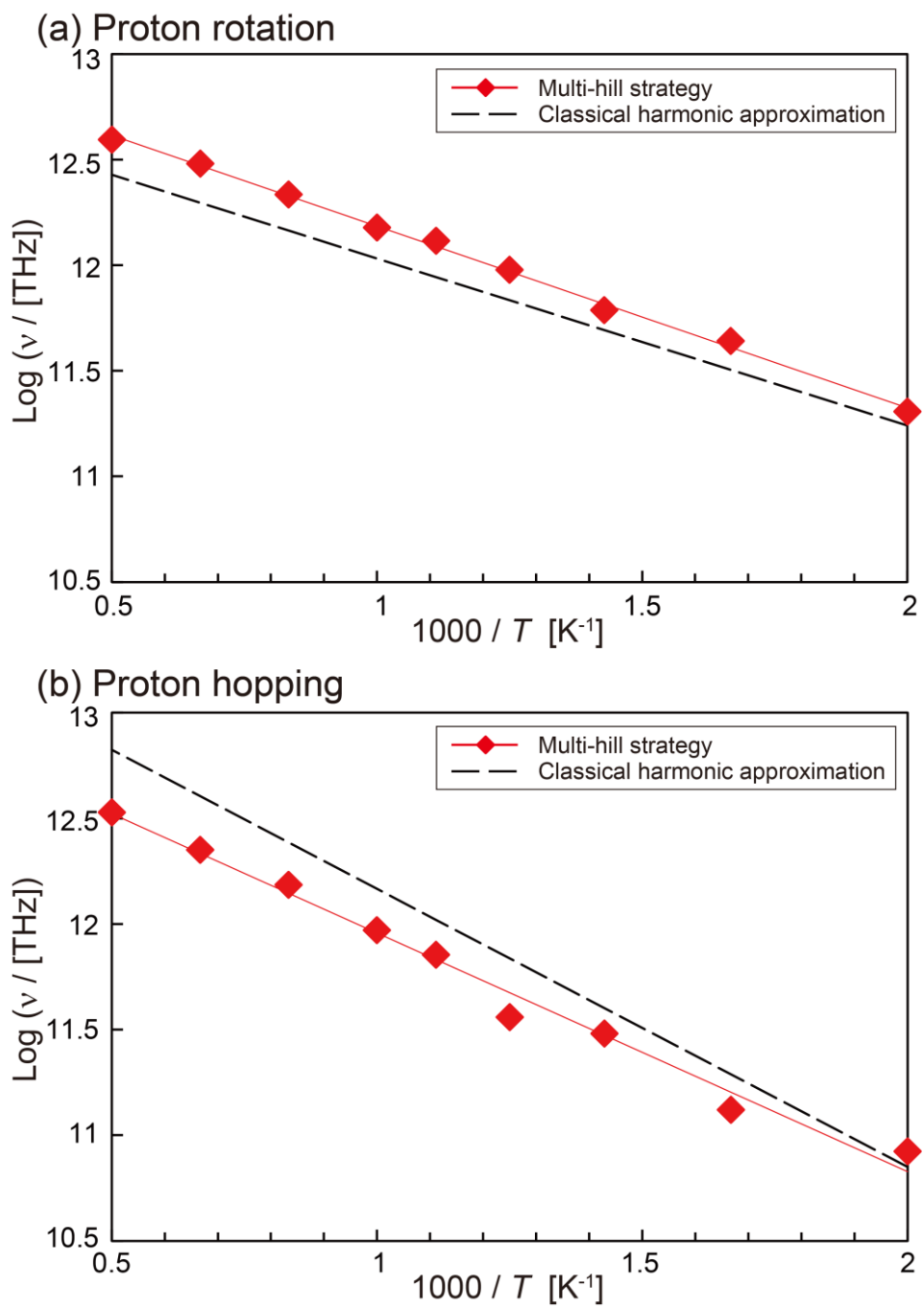

FIGURE 6. Temperature dependence of the estimated jump frequencies for (a) proton rotation and (b) proton hopping under the multi-hill strategy. The estimated jump frequencies under the classical harmonic approximation are also shown for comparison.